\documentclass[12pt]{article}
\usepackage{epsfig,amssymb,amsmath,psfrag,subfigure,rotate,color,wasysym}

\allowdisplaybreaks

\def\XXint#1#2#3{{\setbox0=\hbox{$#1{#2#3}{\int}$ }
\vcenter{\hbox{$#2#3$ }}\kern-.6\wd0}}

\def \be  {\begin{equation}}
\def \ee  {\end{equation}}
\def \ba  {\begin{eqnarray}}
\def \ea  {\end{eqnarray}}
\def \baa {\begin{eqnarray*}}
\def \eaa {\end{eqnarray*}}
\def \lab #1 {\label{#1}}

\newcommand\re[1]{(\ref{#1})}
\def\d{\hbox{{d}\kern-.20em\hbox{l}}}

\def \matrx #1 {\left(\begin{array}{cc} #1 \end{array}\right)}

\newcommand{\Gm}{\Gamma}

\newcommand{\ep}{\epsilon}

\newcommand{\lm}{\lambda}

\begin{document}

\begin{titlepage}

\thispagestyle{empty}

\vspace*{3cm}

\centerline{\large \bf MB Tools reloaded}
\vspace*{1cm}

\centerline{\sc A.V. Belitsky$^a$, A.V.~Smirnov$^b$, V.A.~Smirnov$^c$}

\vspace{15mm}

\centerline{\it $^a$Department of Physics, Arizona State University}
\centerline{\it Tempe, AZ 85287-1504, USA}

\vspace{5mm}

\centerline{\it $^b$Research Computing Center, Moscow State University and}
\centerline{\it Moscow Center for Fundamental and Applied Mathematics}
\centerline{\it 119992 Moscow, Russia}
\vspace{5mm}

\centerline{\it $^c$Skobeltsyn Institute of Nuclear Physics, Moscow State University and}
\centerline{\it Moscow Center for Fundamental and Applied Mathematics}
\centerline{\it 119992 Moscow, Russia}

\vspace{20mm}

\centerline{\bf Abstract}

\vspace{5mm}

We address the problem of evaluation of multiloop Feynman integrals by means of their Mellin-Barnes representation. After a brief overview of available capabilities 
though open source toolkits and their application in various circumstances, we introduce a new code {\tt MBcreate} which allows one to automatically deduce a concise 
Mellin-Barnes representation for a given parametric integral. A thorough discussion of its implementation and use is provided.

\end{titlepage}

\setcounter{footnote} 0


\renewcommand{\thefootnote}{\arabic{footnote}}

\section{Overview}

Feynman parametrization of momentum (or position, for that matter) space integrals is undoubtedly the most widespread tool to perform $D$-dimensional loop integrals  -- see, e.g., recent books
\cite{Smirnov:2012gma,Weinzierl:2022eaz}. After this rather straightforward step, one ends up with an $N$-fold parametric integral of the form
\begin{align}
\label{INdef}
I_N (\{s\})= \int_0^\infty d^N x \, \delta \left( \sum\nolimits_{i=1}^N x_i - 1 \right) \, \mathcal{J} \left(\{x\};\{s\}\right)
\, ,
\end{align}
where the integrand
\begin{align}
\label{JfeynmanIntegrand}
\mathcal{J} \left( \{x\}; \{s\} \right) = 
\mathcal{U}^{p_1} \left( \{x\} \right) \mathcal{F}^{p_2} \left( \{x\}; \{s\} \right)
\prod_{i=1}^N x_i^{a_i-1}
\, ,
\end{align}
is encoded in two Symanzik functions $\mathcal{U} \left(\{x\} \right)$ and $\mathcal{F} \left(\{x\}; \{s\} \right)$, depending on Feynman parameters $\{x\} = \{ x_1, \dots, x_N \}$
and kinematical invariants $\{s\} = \{ s_1, \dots, s_M \}$.  The former are defined by trees and 2-trees of a given Feynman 
graph, respectively. $\mathcal{U}$ is a linear in each $x_i$ polynomial with positive coefficients. The dependence on external kinematical invariants and masses 
$s_k$ enters linearly through the $\mathcal{F}$ polynomial only. In the case of vanishing masses, the latter is linear in each $x_i$ as well but not otherwise. 
These graph polynomials possess, correspondingly, the degrees of homogeneity $h$ and $h+1$ in Feynman parameters, $\mathcal{U} (\{\lambda x\}) = \lambda^h 
\mathcal{U} (\{x\})$ and $\mathcal{F} (\{\lambda x\}; \{s\})= \lambda^{h+1} \mathcal{F} (\{x\}; \{s\})$, where $h$ is the number of loops. Finally, the exponents of 
these polynomials are $p_1=a - (h+1) D/2$ and $p_2 = - a + h D/2$ with $a = \sum_{i = 1}^N a_i$ and individual $a_i$'s corresponding to the powers of appearing propagators 
in the initial momentum integrand. An extensive discussion of their construction can be found in Refs.\ \cite{Bogner:2010kv,Smirnov:2012gma}.

The focus of the present paper is on the calculation of \re{INdef} by transforming it from the real axis to the complex plane where information about integrand's singularities 
will be sufficient to compute $I_N$ making use of powerful theorems of the Complex Analysis. The starting point for this well-known method is based on the following 
Mellin-Barnes (MB) representation
\begin{equation}
\frac{1}{(A+B)^{\lm}} = \frac{1}{\Gm(\lm)}
\int_{\cal C} \frac{d z}{2 \pi i} 
\frac{B^z}{A^{\lm+z}} \Gm(\lm+z) \Gm(-z) \; ,
\label{MB} 
\end{equation}
which allows one to partition a complicated polynomial in terms of its two `simpler' components $A$ and $B$. In this equation, the contour ${\cal C}$ goes from $-i \infty$ to 
$+i \infty$ in the complex plane and the poles of $\Gm(\ldots+z)$ are to its left while the ones of $\Gm(\ldots-z)$ are to its right with these left/right poles corresponding to 
infrared/ultraviolet singularities of the original integral. This formula is usually applied repeatedly enough number of times to a given parametric integral $I_N$ in order to solve 
all $x$-integrations in terms of products of Euler Gamma functions. This yields a sought-after MB representation for a given Feynman integral in the form of an $n$-fold complex 
integral (generally $n \neq N$)
\begin{equation}
 \int_{{\cal C}_1}\ldots \int_{{\cal C}_n}
 \frac{d^n z}{(2 \pi i)^n}
\frac{\prod_i \Gm\left(\alpha_i+\beta_i \ep+\sum_j \gamma_{ij} z_j\right)}
{\prod_i \Gm\left(\alpha'_i+\beta'_i \ep+\sum_j \gamma'_{ij} z_j\right)}
\prod_k s_k^{d_k} 
\label{MBn}
\; .
\end{equation}
The `additive' dependence of the second Symanzik polynomial $\mathcal{F} (\{x\}; \{s\})$ on the kinematical invariants/masses $s_k$ is thus transformed into the 
multiplicative dependence on their $d_k$-powers.
 
The above MB representation (\ref{MBn}) was successfully employed in analytical calculations of Feynman integrals starting with the  seminal work of 
Refs.~\cite{Usyukina:1992jd,Usyukina:1993ch} where  three- and four-point massless ladders at generic values of squared external momenta were 
obtained. Since multiloop Feynman integrals are rather involved objects, they are usually evaluated as a Laurent expansion in the parameter of dimensional 
regularization $\ep=(4-D)/2$, rather than for generic $D$ values. Emerging poles in $\ep$ have different origin reflecting divergent regions in the initial momentum 
integrals: they can stem from ultraviolet, infrared, collinear etc. domains. A systematic study of MB representation (\ref{MBn}) for dimensionally regularized Feynman 
integrals was initiated in Refs.\ \cite{Smirnov:1999gc,Tausk:1999vh} where two complementary strategies for resolving occurring singularities in $\ep$ near $\ep=0$ 
were devised. More than that, two public computer codes based on these techniques were developed in \cite{Smirnov:2009up,Czakon:2005rk}, respectively. This 
laid out the foundation for a widespread use of the MB techniques by QFT practitioners, see, e.g., Chapter~5 of~\cite{Smirnov:2012gma} for a review.

Admittedly the MB method had seen its better days in the rear-view mirror as it passed its pinnacle on the stage of calculation tools being superseded by the introduction 
of canonical integral bases~\cite{Henn:2013pwa} within the method of differential equations (DEs) \cite{Kotikov:1990kg,Gehrmann:1999as}. A historical remark is in order to 
make this point clear to the reader. To reveal the so-called BDS Ansatz~\cite{Bern:2005iz} for four-gluon scattering amplitudes at three-loop order, it was necessary to 
evaluate two four-leg Feynman integrals associated with triple ladder-box and tennis-court graphs. This was performed in~\cite{Smirnov:2003vi} and~\cite{Bern:2005iz}, 
respectively, making use of the MB technique at the time when no computer codes were yet available thus making the task colossally tedious. Application of the same 
approach to all master integrals of the above two families of Feynman integrals would not be feasible, but with the use of DEs for canonical bases of integrals this goal can 
successfully be achieved as was shown in Ref.~\cite{Henn:2013tua}. More than that, in a similar manner, master integrals for all four-leg massless on-shell non-planar 
graphs were also evaluated \cite{Henn:2016jdu,Henn:2020lye}.

Nevertheless, the MB method remains powerful enough to keep its runner-up position and can be applied in conjunction with DEs in order to fix their boundary conditions, 
see, e.g., Refs.\ \cite{Henn:2014lfa,Caola:2014lpa}, or, in certain circumstances, it is the only available choice when DEs cannot be used or face their own vices. A particularly 
suitable niche for the application of the MB technique is in the analysis of asymptotic behavior of Feynman graphs for small/large values of occurring kinematical invariants/masses 
$s_k$: like Sudakov and heavy mass limits, just to name a few. Leading contributions in these cases are revealed with the help of a strategy known as the Expansion by  
Regions~\cite{Beneke:1997zp} (see also \cite{Smirnov:1999bza,Smirnov:2002pj,Smirnov:2012gma}). This is accomplished by applying the public Mathematica code {\tt asy} 
\cite{Pak:2010pt,Jantzen:2012mw},--- also available as the {\tt SDExpandAsy} command with the {\tt FIESTA5} distribution package \cite{Smirnov:2021rhf},--- which is based 
on the analysis of the geometry of 
polytopes associated with the two Symanzik polynomials $\mathcal{U} (\{x\})$ and $\mathcal{F} (\{x\}; \{s\})$. It determines all leading contributions to the $I_N$ 
integral by scanning over various scaling behaviors of the Feynman parameters with asymptotic values of kinematical invariants. The output is given as parametric 
integrals of the $I_N$ type but with reduced, scale-independent Symanzik polynomials, $\widetilde{\mathcal{U}} (\{x\})$ and $\widetilde{\mathcal{F}} (\{x\})$. Since 
there is no dependence on kinematical variables left, DEs are powerless and the MB approach is the only game in town. This strategy was recently applied on different occasions, 
see \cite{Belitsky:2021huz,Belitsky:2022itf}, which compelled updates to existing routines of the MB toolbox as well as development of a new code, which will be described below.

The subsequent presentation is organized as follows.  Sect.~\ref{MBcreateSection} describes the main contribution of this work through the code {\tt MBcreate}, which generates 
a concise MB representation for a given Feynman integral. Next, Sect.~\ref{MBtoolsSection} provides an exposition of existing codes connected with the MB representations, 
which allows one to solve MB integrals in the form of the Laurent expansion in $\ep$ with analytic coefficients expressed in terms of Riemann zeta values. Conclusions with 
several appendices culminate the paper.
 
\section{Introducing {\tt MBcreate.m}}  
\label{MBcreateSection}

The first of order of business on the way to apply available MB tools is to derive an optimal MB representation \re{MBn} for a given Feynman-parameter integral \re{INdef} with a 
minimal number $n_0$ of complex integrations. For generic momentum-space integrals, one can proceed in two different ways: construct Feynman parametric representation
for the entire multiloop integrand, then deduce corresponding MB integrals, or do it loop-by-loop, i.e., derive an MB representation for a one-loop subintegral, then embed it into a 
larger two-loop integral and so on. It turns out that the global route does not yield the minimal value for $n_0$. An example to this point is the very first analytical calculation of 
dimensionally regularized double boxes \cite{Smirnov:1999gc} where the global parametric representation produced five MB integrations but later it was 
observed~\cite{Anastasiou:2000kp} that in the loop-by-loop approach that it reduces down to four. The latter method was then successfully used in planar-graph calculations, 
e.g., \cite{Smirnov:2003vi,Bern:2005iz}. It found its automatic implementation in the public code {\tt AMBRE.m}~\cite{Gluza:2007rt}. Yet another algorithm to derive an optimal MB 
representation was proposed in Ref.~\cite{Prausa:2017frh}, however, its computer implementations is not available so far. 

Starting with a generic Feynman integral \re{INdef} depending on $M$ kinematical variables $s_k$, it is only natural to isolate them first in a factorized form \re{MBn} by means of 
the repeated use of Eq.\ \re{MB} at the cost of introducing $M-1$ MB integrations. The leftover is then a product of several Feynman-parameter polynomials $F_j$ (with positive 
coefficients) raised to certain (generally) complex powers. Similarly, as discussed in the previous section, the application of expansion by regions yields $s_k$-independent reduced 
Symanzik polynomials in parametric integrands. In either case, one has to construct an optimal MB representation for these. One can of course proceed by trial and error on a 
case-by-case basis looking for the magic number $n_0$. This was done in Refs.\ \cite{Belitsky:2021huz,Belitsky:2022itf}. However, this is extremely time-consuming. In the lack of 
a proof of what a numerical value $n_0$ might be a priory, a routine that can search for its optimal value needs to be developed to tackle this problem. So the lowest value of $n_0$ 
that it finds will constitute an efficiency criterion.

Therefore, consider a Feynman parametric integral independent of kinematical invariants
\begin{equation}
\int_0^\infty d^N x \prod_i x_i^{a_i-1}  \, \delta \left( \sum\nolimits_{i=1}^N x_i - 1 \right)  \prod_j F_j^{p_j} \left(\{x\}\right)
\label{ParInt}
\, .
\end{equation}
Here $F_j$ are polynomials with positive coefficients linear in each $x_i$, raised to powers $ p_j = b_j \ep+c_j$; $a_j$ are integers, while $b_j,c_j$ are rational numbers when 
these are thought of as outputs of the Expansion by Regions\footnote{\label{FootnoteUF} In this case, it is obvious that $F_1 = \widetilde{\mathcal{U}}$ and $F_2 = 
\widetilde{\mathcal{F}}$.}, or complex when it is a result of kinematical split-up alluded to at the top of the previous paragraph. Notice that $a_i$'s can also be considered generally 
complex-valued if an auxiliary analytic regularization is imposed. 
This latter setup is particularly relevant for initially finite parametric integrals where one can choose to set the number of space-time dimensions down to four, i.e., $\ep=0$. However, 
since its asymptotic expansion with expansion by regions generates individually divergent contributions an intermediate regularization is nevertheless required. It has to be 
imposed however in a manner that does not violate the rescaling invariance of the original parametric integral under $\{x\} \to \{\lambda x\}$ transformation. The preservation 
of this property is crucial for maintaining the opportunity to apply the so-called Cheng-Wu theorem \cite{ChengWu} to the above integral. For reader's convenience and completeness 
of this presentation, the theorem is reviewed in Appendix \ref{ChengWuTheoremAppendix} and boils down to reducing the delta-function constraint down to a smaller subset of 
Feynman parameters. A particularly convenient choice is $\delta\left(x_{i_0}-1\right)$ for a single ad hoc $i_0$. In certain calculations, one eliminates the delta function constraint 
first in favor of symmetric treatment of all integrals involved, be it Feynman-parameter or proper-time integrals, see, e.g., \cite{Belitsky:2021huz}. The procedure devised below is 
applicable to those circumstances as well.

The procedure is built on the following two transformations: integration over an $x$-parameter, if possible, making use of the integral
\begin{align}
&\int_{0}^{\infty} dx \, x^p (a x+b)^q
=\frac{\Gamma (p+1) \Gamma (-p-q-1)}{\Gamma (-q)}a^{-p-1} b^{p+q+1} \,,
\label{TabInt1}
\end{align}
introduction of an MB integration \re{MB} in order to apply \re{TabInt1}. Obviously, the first $x$-integrations which have to be performed in Eq.\ \re{ParInt} are the ones over 
non-overlapping subsets of variables defining the $F_j$ polynomials. Without loss of generality it suffices to address the case of just two polynomials in the integrand $F_1$ 
and $F_2$ (see footnote \ref{FootnoteUF}). Suppose that there are several 
variables with this property, i.e., $F_2$ depends on all of the Feynman parameters while $F_1$ is independent of a subset $\Sigma = \{ x_\ell \}$ of these\footnote{This is a typical 
situation for a bulk of contributions stemming from expansion by regions.}. Then $F_2=F_{2,1} x_\ell+F_{2,0}$ and by means of \re{TabInt1}, we obtain the product $F_{j,1}^{p_1} 
F_{j,0}^{p_0}$. Then one repeats this step for the next variable from $\Sigma$ provided it belongs to either $F_{j,1}^{p_1}$ or $F_{j,0}^{p_0}$ but not both. After such integrations 
become impossible, one is forced to use the MB partition \re{MB} first before applying \re{TabInt1} again. It is at this step that an optimal choice of the the decomposition of the
progenitor polynomial $F_{j,\dots}$ into its simpler components becomes crucial for the most efficient MB representation. The key question is to minimize the number $n_0$ of those 
complex integrations. 

The generic steps outlined above were implemented in the Mathematica package {\tt MBcreate.m}, which attempts to minimize the value of $n_0$. In particular, {\tt MBcreate.m} 
examines and applies the following procedures one-by-one, not necessarily in the order listed, unless it is explicitly specified.

\begin{itemize}
\item Factorizes kinematic invariants $s_k$ from $F_j$'s first: if $F_j = f_{j,0} + s_k f_{j,1}$, an MB representation \re{MB} is introduced to split up $f_{j,0}$ and $s_k f_{j,1}$.
\item Implements the change of variables  $x_i = \eta  \xi, x_j = \eta  (1 - \xi)$ for two Feynman parameters entering integrands and obeying the conditions: (i) the dependence of each of 
the functions $F_j$ of $\eta$ is at most linear, (ii) no more that two $F_j$'s depend on it. Otherwise, introduces an MB representation, integrates with respect to $\eta$. Next 
introduces yet another MB decomposition \re{MB} with subsequent integration over $\xi$.
\item Searches through all $\{x_i,x_j\}$ pairs and find cases where only one of the $F_j$ function depends on a single variable, say $x_i$, not the sum of the two $x_i + x_j$. Splits 
up that function into two terms, one depending on the sum and the rest, solves the resulting integration with Eq.~\re{TabInt1}.
\item Tries all decompositions of the form $F_j = x_i  F_{j,1} + F_{j,0}$, where $F_{j,0}$ does not depend on $x_i$, with both $F_{j,1}$ and $F_{j,0}$ being factored into monomials 
accompanying residual polynomials. Splits $F_j$ by introducing an MB integration.
\item Scans all decompositions $F_j = x_i  F_{j,1} + F_{j,0}$, where $F_{j,0}$ is $x_i$-independent and splits them up with the MB representation \re{MB} provided $F_{j,1}$ and/or $F_{j,0}$
already exist in the list of functions populating the integrand. This reduces the number of polynomials which could potentially yield a higher value of $n_0$.
\item Searches for possible splitting based on the form $F_j = x_i  F_{j, 1} + F_i$ where $F_i$ is one of the factors already present in the integrand.
\item Tries the decompositions $F_j = x_i  F_{j,1} + F_{j,0}$ where $ F_{j,0}$ can depend on $x_i$ but is factorized into a product of simpler, lower-degree polynomials.
\item Searches for `similar' functions $F_i$ and $F_j$ defined by the condition that $F_i - F_j$ is given by a difference of two monomials. Splits one them, say $F_i$, into the monomial 
associated with it and the rest, even more cognate with $F_j$ function, by means of Eq.\ \re{MB}.
\item If none of the above procedures meet their requirements, chooses a Feynman variable $x_i$, introduces MB representations for all $F_j$'s in the integrand but one
and performs the integration over $x_i$ of the last remaining polynomial with the help of Eq.\ \re{TabInt1}.
\end{itemize}
{\tt MBcreate.m} automatically applies all of the above strategies and then solves the resulting parameter integrals whether they require an MB representation or not. The output
is given by the product of ratios of Euler Gamma functions as in the integrand of Eq.\ \re{MBn}.

On extremely rare occasions, when an additional regularization is called for successful resolution of singularities discussed in the next section, an output can be encountered
with Euler Gamma of arguments depending on the parameter of analytic regularization only. These have to be scrapped and redone by manually reshuffling the indices of $x$'s.

\section{MB tools overhauled}  
\label{MBtoolsSection}

Having derived an MB representation \re{MBn} for a parametric integral, one has to solve it as a Laurent series in $\ep$ up to a desired order, with coefficients which are given 
by MB integrals independent of $\ep$, i.e., pure numbers. To this end, it is necessary to resolve the singularity structure in $\ep$. As was already addressed in the introductory 
section, one can use either {\tt MB.m} or {\tt MBresolve.m} for that purpose, which were delivered in Refs.\ \cite{Czakon:2005rk,Smirnov:2009up}, respectively. The initial point of 
{\tt MB.m} is to apply {\tt MBoptimizedRules} command in order to find straight contours and values of $\ep$ obeying the rules for the contour choice formulated immediately after 
Eq.\ \re{MB}. Such contours do not always exist from the get-go\footnote{In the latest calculations from~\cite{Belitsky:2021huz,Belitsky:2022itf} they do not exist as a rule.}. To 
alleviate the problem one can introduce an auxiliary analytic regularization complementary to the dimensional one and then proceed to contour determination with this command. 
There is no universal prescriptions how to do this in a systematic way and it is not straightforward. Due to these complications, it is preferable to rely on {\tt MBresolve.m} instead. 
As it was explained in detail in Ref.\ \cite{Smirnov:2009up}, the code searches for optimal straight contours for the resolution of singularities in $\ep$. Only on rare occasions, the
code is unable to perform and this calls for an auxiliary analytic regularization. A recommended way of doing it in a systematic fashion is to provide additive terms to all $a_i$ in Eq.\ 
\re{ParInt} proportional to a parameter, say, $\lm$, i.e., $a_i \to a_i + r_i \lambda$ with the total sum $\sum_i r_i$ equal to zero\footnote{The last condition is important because it 
does not affect re-parametrization invariance of parametric integrands to choose a ``gauge" condition on one of the $x$'s. Also numerical checks of Laurent expansions with 
{\tt SDExpandAsy} command of {\tt FIESTA} can be used provided this condition is fulfilled.}. The second reason why it is advantageous to apply {\tt MBresolve.m} rather than 
{\tt MB.m} is that {\tt MBresolve.m} is much faster and this turns out rather important if the number of MB integrations is very large.  A comment is in order about these two codes: 
every so often they used to produce real shifts of the contour which were integers this yielded error messages in the subsequent steps of analyses. The matter was resolved by 
increasing the precision\footnote{If this obstruction still persists, a user can further increase the precision by opening the packages  {\tt MB.m} and {\tt MBresolve.m}, searching for 
{\tt Rationalize} command and making the change by her/him-self.} of the conversion of decimal to rational output to $10^{-5}$.

The next step is to evaluate pure number MB integrals involved.  This is accomplished by running the command {\tt DoAllBarnes} from {\tt barnesroutines.m} \cite{KosowerBarnes} 
which automatically applies the first and the second Barnes lemmas and thereby performs some integrations in terms of Euler Gamma functions. The current version of the routine 
does not include a plethora of corollaries of the lemmas and they have to be applied by hand as in recent studies \cite{Belitsky:2022itf}. A case in point is 
\begin{align}
\label{Barnes2-1} 
&
\int_{\cal C} \, \frac{dz}{2 \pi i}
\frac{\Gamma (a+z) \Gamma (-b-z) \Gamma (b+z) \Gamma (d-z)}{z}
=-\Gamma (2-a) \Gamma (a) \Gamma (-b) \Gamma (b)
\, \nonumber \\
&\qquad\qquad +\Gamma (2-a) \Gamma (-b) \Gamma (a-b-1) \Gamma (b-a+2)
\frac{\psi(1-a)-\psi(-b)}{\Gamma (1-a)
   \Gamma (1-b)} \nonumber \\
&\qquad\qquad -\frac{1}{b^2}\Gamma (a-b) \Gamma (b-a+2) (b (\psi(b-a+2)+\gamma_{\rm E} )-1)    
\, ,
\end{align}
where the pole $z=0$ stays to the left of an integration contour and the pole $z=-b$ positioned to the right of the integration contour. A very long comprehensive list of similar
formulas is provided in the Appendix \ref{CorollariesAppendix} as an attachment.

An alternative and a much faster route in many circumstances, however, is immediately after the application of  {\tt DoAllBarnes} to bypass the use of corollaries of Barnes lemmas and 
turn to numerical analyses of remaining MB integrals. In practice, the computation of the latter is not problematic since these converge very well at large imaginary values of 
$z$-integration variables due of the exponential suppression stemming from Euler Gamma  functions involved. Thus, calculating these with sufficiently high precision is possible. 
Then one can use the {\tt PSLQ} algorithm~\cite{PSLQ:1999} to obtain analytic results provided a basis of numbers, typically values of Riemann zeta function, entering the 
final result is known. For one-dimensional MB integrals, the current version of {\tt NIntegrate} with {\tt GlobalAdaptive} strategy in {\tt Mathematica} can achieve the precision of 
100 or more with ease and then the built-in command {\tt FindIntegerNullVector} allows one to successfully recognize transcendentals\footnote{A code implementing it is provided 
in Appendix \ref{PSLQAppendix}.}. Currently, {\tt Mathematica} cannot handle well $n$-fold integrals for $n>2$ with sufficiently high precision and this sets a strong limitation of this 
calculational strategy. For instance, for two-fold MB integrals, only an older version of {\tt Mathematica}, e.g., v.5.2, permits one to gain sufficient precision (topping at 40) with the 
{\tt DoubleExponential} option for {\tt NIntegrate}. A lower available precision imposes a very strong restriction on the dimension of the basis of transcendental numbers. After analytic 
results have been found with {\tt PSLQ}, it is advisable to use {\tt MBintegrate} of {\tt FIESTA}  to the intermediate output of {\tt MBresolve} to verify the former numerically.

While the original {\tt MB} package was distributed via hepforge {\tt https://mbtools.hepforge.org/}, the current development of {\tt MBcreate.m} is undergone with the use of {\tt git} 
and {\tt bitbucket}. Both {\tt MBcreate.m} and all other {\tt MB} codes are collected in the same repository and can be freely downloaded from there: 
{\tt https://bitbucket.org/feynmanIntegrals/mb/src/master/}.

Most of the codes require {\tt Mathematica} and simply work when loaded there. However, the {\tt MBintegrate} command performs integration with the use of {\tt fortran} generated 
codes and requires the {\tt gfortran} compiler (which can normally be installed with package managers such as {\tt apt-get}) as well as some libraries. While the Cuba integration 
library \cite{CUBA} by T.~Hahn is shipped with the package and works perfectly with modern compilers, the original {\tt MB} code used also the cernlib library for the evaluation of 
polylogarithms, but the cernlib is no longer supported, and there might be a problem to install it at modern computers. Hence a code based on a small portion of cernlib which was 
provided by M.~Czakon is also included in the repository. All libraries can be compiled by calling {\tt make} in the package folder.

The new package is also accompanying this submission as an ancillary file,  for reader's convenience, along with a Mathematica notebook with a thoroughly worked out 
example {\tt example.nb}.

\section{Conclusion}

This work introduced a new package for the conversion of Feynman integrals into an MB form with a minimal number of complex integrations, {\tt MBcreate}. Also an update to several
routines in the MB toolbox was provided to have error-free outputs at each step of analytical calculation of Laurent expansion of Feynman integrals. These were thoroughly tested
against calculations done mostly ``by hand" in Refs.\ \cite{Belitsky:2021huz,Belitsky:2022itf}.

For completeness, it is worth pointing out that while the strategy outlined above heavily relies on the {\tt PSLQ} algorithm, there is yet another alternative way to evaluate an MB integrals 
by transforming them into infinite series representation by closing integration contours and taking residues with a help of computer code presented in Ref.~\cite{Ochman:2015fho}. The 
very problem of finding series representations for a given MB integrals was recently analyzed in \cite{Ananthanarayan:2020fhl,Ananthanarayan:2021not} making use of an approach based 
on conic hulls. A public computer code was also given there.

\section*{Acknowledgments}

We would like to thank M.~Czakon for providing the codes required to make {\tt MBintegrate} to work on modern computers. The work of A.B.\ was supported by the U.S.\ National 
Science Foundation under the grant  No.\ PHY-2207138. The work of  A.S and V.S.\  was supported by the Russian Science Foundation under the agreement no. 21-71-30003 
(application of MB representations to research in elementary particle physics) and by the Ministry of Education and Science of the Russian Federation as part of the program of the 
Moscow Center for Fundamental and Applied Mathematics under Agreement No.\ 075-15-2019-1621 (development of algorithms creating MB representations).

\appendix

\section{Folklore Cheng-Wu theorem}
\label{ChengWuTheoremAppendix}

Consider an $N$-fold integral on the standard simplex
\begin{align}
\label{iniIN}
I_N = \int_{\mathcal{S}_x} d^N x \, J (\{x\})
\, , \qquad
\mathcal{S}_x = \left\{ \{x\} \in \mathbb{R}^N:\sum\nolimits_{i = 1}^N x_i = 1, x_i \geq 1 \right\}
\, ,
\end{align}
with the integrand being a homogeneous function of the $x_i$ variables of degree $r$, i.e., $F (\{\lambda x\}) = \lambda^r F (\{x\})$. This integral is not invariant under 
this rescaling instead it has the degree $r + N$. To alleviate this predicament, perform a projective transformation by passing to another set of variables $\{y\}$ as 
\begin{align}
x_i = y_i/\left( \sum\nolimits_{i=1}^N y_i \right)
\, , \qquad 
i = 1, \dots, N
\, .
\end{align}
This change leaves the simplex domain invariant $\mathcal{S}_x = \mathcal{S}_{y}$. Then, taking into account the emerging Jacobian
\begin{align}
d^N x = d^N y/\left( \sum\nolimits_{i=1}^N y_i \right)^N
\, ,
\end{align}
the integral in these new variables becomes
\begin{align}
I_N = \int_{\mathcal{S}_y} d^Ny \, J (\{y\})/\left( \sum\nolimits_{i=1}^N y_i \right)^{r + N}
\, ,
\end{align}
and is explicitly rescaling invariant.

The latter property becomes crucial in the efficient solution of integrals by means of the application of the so-called Cheng-Wu theorem \cite{ChengWu}, which specifies 
various possible choices for multifold integrations. To formulate it, introduce a constraint on the integration variables in the integrand but integrate over unconstrained 
$\mathbb{R}_+^N$ space, i.e., 
\begin{align}
\label{Iinvariant}
I_N = \int_0^\infty d^N x \, \delta \left( \sum\nolimits_{i=1}^N x_i - 1 \right) \, \mathcal{J} (\{x\})
\quad \mbox{with} \quad
\mathcal{J} (\{x\})
=
J (\{x\})/\left( \sum\nolimits_{i=1}^N x_i \right)^{r + N}
\, .
\end{align}
The Cheng-Wu theorem states that one can freely change the argument of the Dirac delta function to
\begin{align}
\label{SubsetDomain}
\sum\nolimits_{i=1}^N x_i - 1 \to \sum\nolimits_{i \in \Sigma} x_i - 1
\end{align}
with $\Sigma$ being a subset of $N$ labels. In particular, one can choose just one, say $i_0$, in which case this variable is set to $x_{i_0} = 1$ and the 
unconstrained integration is performed over the remaining $N-1$ ones. It is important to realize that one could not have applied it to the original integral \re{iniIN} 
since it is not rescaling invariant.

Though proofs of the Cheng-Wu theorem can be found in Refs.\ \cite{Grozin,Smirnov:2012gma,Weinzierl:2022eaz}, it is enlightening however to present it again. One can 
directly prove it making use of the Stokes's theorem \cite{Weinzierl:2022eaz} applied to \re{Iinvariant}, but it is instructive to invoke instead its relation to the Feynman parameter 
integral for a graph as given in Eq.\ \re{JfeynmanIntegrand}. Using it as a starting point, one can next integrate-in a variable to obtain the well-known Schwinger, aka alpha, 
representation for the integral
\begin{align}
\label{SchwingerRep}
I_N = 
\frac{1}{\Gamma(a - h D/2)}
 \int_0^\infty d^N \alpha \, \left[\mathcal{U} \left(\{\alpha\} \right) \right]^{-D/2}
\exp\left( - \frac{\mathcal{F}  \left(\{\alpha\} \right)}{\mathcal{U}  \left(\{\alpha\} \right)}\right)  \prod_{i=1}^N \alpha_i^{a_i-1}
\, .
\end{align}
Obviously, the integrand \re{JfeynmanIntegrand} is obtained from this one by the rescaling $\{\alpha\} = \rho \{x\}$ and subsequent integration with respect to $\rho$.
Using this representation as a starting point, the proof of the Cheng-Wu theorem becomes elementary. Namely, resolve the unity in terms of a constraint involving
a subset of $\alpha$'s, as on the right hand side of Eq.\ \re{SubsetDomain}
\begin{align}
1 = \int_0^\infty d \sigma \, \delta \left(  \sum\nolimits_{i \in \Sigma} \alpha_i - \sigma \right)
\, ,
\end{align}
and substitute it into the integrand of Eq.\ \re{SchwingerRep}. Next, change the variables to $\{\alpha\} = \sigma \{x\}$ and use scaling properties of all functions 
involved to get the integral
\begin{align}
I_N = 
\frac{1}{\Gamma(a - h D/2)}
\int_0^\infty d^N x\, 
&
\prod_{i=1}^N x_i^{a_i-1}
\,
\delta \left(  \sum_{i \in \Sigma} x_i - 1 \right)
\nonumber\\
& \times [\mathcal{U} (\{x\})]^{-D/2}
\int_0^\infty d \sigma \, \sigma^{a - h D/2-1} \exp\left( - \sigma \frac{\mathcal{F} (\{x\})}{\mathcal{U} (\{x\})}\right)  
\, .
\end{align}
Finally integrating over $\sigma$ gives original integrand with a constraint encompassing only a subset of integration variables. 

\section{Corollaries of Barnes lemmas (V.S. 2004)}
\label{CorollariesAppendix}

For readers convenience, a very long list of corollaries of Barnes lemmas by V.S.\ is attached with this paper in the file {\tt barnes.txt}.

\section{PSLQ}
\label{PSLQAppendix}

Since the original Broadhurst's PSLQ code is not freely available to general public, here is a `one-line' routine based on the {\tt FindIntegerNullVector} 
command in Mathematica:
\begin{verbatim}
PSLQ[num_?NumericQ, basis_?VectorQ] := 
Module[{coefficients, result}, 
coefficients = FindIntegerNullVector[Prepend[N[basis, Precision[num]], num]];
result = Rest[coefficients].basis/First[coefficients]; 
Sign[N[result]] Sign[num] result];
\end{verbatim}
The syntax is self-explanatory from the following example:
\begin{verbatim}
In[1]:= 
Num = -4.2306193701686518817682268282580510275171911584045617944546633\
2782312410899782814554047544567313363330613025333597361955278613729766\
333`99.43031384975686;

Basis = {1, EulerGamma, EulerGamma^2, EulerGamma^3, EulerGamma^4, Pi^2, 
Pi^4, EulerGamma Pi^2, EulerGamma^2 Pi^2, EulerGamma^3 Pi^2, 
EulerGamma Pi^4, Zeta[3], Zeta[3] EulerGamma};

PSLQ[Num, Basis]

Out[1]= 1/16 (-76 - 44 EulerGamma - 22 EulerGamma^2 - \[Pi]^2 
+ 14 EulerGamma \[Pi]^2 - 24 Zeta[3])
\end{verbatim}
Above, an overcomplete basis is used for demonstration purposes only of the uniqueness of the reconstruction. A more efficient choice would require lower precision of 
numerical inputs.


\end{document}